\documentclass[twocolumn,aps,prc,showpacs,preprintnumbers]{revtex4}%
\usepackage{amssymb}
\usepackage{amsmath}
\usepackage{amsfonts}
\usepackage{graphicx}
\usepackage{dcolumn}
\usepackage{bm}
\begin{document}
\title{Non-inertial effects in reactions of astrophysical interest}
\author{C.~A.~Bertulani,  J.T.~Huang, and P.~G.~Krastev\footnote{Current address:
San Diego State University, Department of Physics, 5500 Campanile
Dr., San Diego, CA 92182-1233}}
\address{Department of Physics, Texas A\&M University, Commerce,
TX 75429, USA}\email[]{e-mail: carlos_bertulani@tamu-commerce.edu,
jhuang1@leo.tamu-commerce.edu, pkrastev@sciences.sdsu.edu}

\begin{abstract}
We discuss the effects of non-inertial motion in reactions occurring
in laboratory, stars, and elsewhere. It is demonstrated that
non-inertial effects due to large accelerations during nuclear
collisions might have appreciable effects nuclear and atomic
transitions. We also explore the magnitude of the corrections
induced by strong gravitational fields on nuclear reactions in
massive, compact stars, and the neighborhood of black holes.

\end{abstract}
\date{\today}

 \maketitle

\section{Introduction}

Extremely large accelerations occur when atomic nuclei collide. For
instance, two lead nuclei in a head on collision with a center of
mass kinetic energy of 500 MeV, reach a closest distance of 19.4 fm
before they bounce back and move outward. At this distance each
nucleus accelerates with an intriguing $\sim 10^{27}$ m/s$^2$. Very
few other physical situations in the Universe involve nuclei
undergoing such large accelerations, usually related to
astrophysical objects, as in the vicinity of neutron stars and black
holes, where huge gravitational fields exist. In this article we
explore the effects of large accelerations and large gravitational
fields,  and their possible influence on nuclear reactions in the
laboratory and in astrophysical environments. Nuclear reactions are
crucial for the formation of stellar structures and their rates
could be affected by various factors. To our knowledge, the effect
of large gravitational fields on nuclear reaction rates in stars has
not been considered so far.

As mentioned in the previous paragraph, atomic and nuclear systems
undergo large accelerations during reactions. The effect of
acceleration is observed in terms of excitations followed by decay
of these systems. If we consider two-body reactions, there are two
systems of reference which are often used to describe the effects of
the collision: (a) the center-of-mass (cm) system of the two nuclei
and (b) the system of reference of the excited nucleus. System (b)
is appropriate to use when the intrinsic properties of the excited
nucleus is described in some nuclear model. A typical example is the
case of Coulomb excitation. One assumes that the nuclei scatter and
their cm wave functions are described by Coulomb waves due to the
Coulomb repulsion between the nuclei. Then one considers the
residual effect of the Coulomb potential on the motion of the
nucleons inside the nuclei. This is done by expanding the  Coulomb
potential in multipoles and using the high order terms (higher than
first order) as a source of the excitation process. In this approach
one illustrates the privileged role of the cm of the nuclear system:
the net effect of the external forces is to (i) accelerate all the
particles together, along with the cm of the system, and (ii) to
change the intrinsic quantum state of the system through the spatial
variation of the interaction within the system. Thus the theoretical
treatment of accelerated many-body systems is well under control in
non-relativistic dynamics.

In the non-relativistic case, the separation of variables into
intrinsic motion and relative motion between the cm of each nucleus
is a simple algebraic procedure. A problem arises when one wants to
extend the method to describe intrinsic excitations of relativistic
many-body systems.  Very few works exist in the literature
addressing this problem. The reason is that for nuclear reactions in
the laboratory, the effect is expected to be very small, a common
belief which must be tested. Another other reason is that in stellar
environments where the gravitational fields are large, huge
pressures develop, "crushing" atoms, stripping them from their
electrons, and ultimately making nuclei dissolve into their
constituents. Effects of nuclear excitation are not relevant in the
process. But, on the other hand, nuclear reactions are crucial for
the formation of stellar structures and their rates could be
affected by minor effects such as those explored in this article.

Nuclei participating in nuclear reactions in a gaseous phase of a
star follow inertial trajectories between collisions with other
nuclei. Such trajectories are ``free fall'' trajectories in which
all particles within the nucleus have the same acceleration. That is
surely true in the non-relativistic case, but not in the
relativistic one because retardation effects lead to corrections due
to the nuclear sizes. The central problem here is the question
regarding the definition of the center of mass of a relativistic
many body system. We have explored the literature of this subject
and found few cases in which this problem is discussed. Based on
their analysis we show that relativistic effects introduce small
corrections in the Lagrangian of a many-body system involving the
magnitude of their acceleration. We follow Refs.
\cite{Landau,Pryce48,fewell}, with few modifications, to show that a
correction term proportional to the square of the acceleration
appears in the frame of reference of the accelerated system. To test
the relevance of these corrections, we make a series of applications
to nuclear and atomic systems under large accelerations.

\section{Hamiltonian of an accelerated many-body system}

Starting with a Lagrangian of a free particle in an inertial frame
and introducing a coordinate transformation into an accelerated
frame  with acceleration $\mathcal{A}$, a ``fictitious force" term
appears in the Lagrangian when written in coordinates fixed to the
accelerated frame. Thus, in an accelerated system the Lagrangian $L$
for a free particle can be augmented by a (non-relativistic)
interaction term of the form $-m\mathcal{A}z$, that is
\begin{equation}
L=-mc^{2}+\frac{1}{2} mv^{2}-m\mathcal{A}z, \label{Lnr}
\end{equation}
where $z$
is the particle's coordinate along the direction of acceleration of
the reference frame~\cite{Landau}.

In the relativistic case, the first step to obtain the Lagrangian of
a many body system in an accelerated frame is to setup an
appropriate measure of space-time in the accelerated frame, i.e. one
needs to find out the proper space-time metric. The free-particle
action $S=-mc\int ds$ requires that $ds=(c-v^2/2c+{\cal A}z)dt$,
which can be used to obtain $ds^2$. To lowest order in $1/c^2$ one
gets
\begin{equation}
ds^{2}=c^{2}\left(1+\frac{\mathcal{A}z}{c^{2}}\right)^{2}dt^{2}
-dx^{2}-dy^{2}-dz^{2}=g_{\mu \nu}d\xi^{\nu}d\xi^{\mu},
\label{metr1}%
\end{equation}
where ${\bf v}dt = d{\bf r}$ was used, with $d\xi^{\mu}=(cdt,dx,dy,dz)$ and
$g_{\mu\nu}=\left(  g_{00}%
,-1,-1,-1\right)  $, $g_{00}=\left( 1+\mathcal{A}z/c^{2}\right)
^{2}$. The indices $\mu$ run from $0$ to $3$. Eq.~(\ref{metr1})
gives a general form for the metric in an accelerated system. This
approach can be found in standard textbooks (see, e.g. ref.
\cite{Landau}, $\S$ 87).

From the definition for the Hamiltonian,
$H=\mathbf{p}\cdot\mathbf{v}-L$, with ${\bf p}=\partial
L/\partial\mathbf{v}=m{\bf v}/\sqrt{g_{00}-v^2/c^2}$, and using the
action with the metric of Eq.~(\ref{metr1}), after a straightforward
algebra one finds
\begin{equation}
H=\frac{g_{00}mc^{2}}{\sqrt{g_{00}-\frac{v^{2}}{c^{2}}}}=c\sqrt{g_{00}\left(
p^{2}+m^{2}c^{2}\right)  }. \label{ham1}%
\end{equation}
Expanding $H$ in powers of $1/c^{2}$, one obtains
\begin{equation}
H    =\frac{p^{2}}{2m}\left(  1-\frac{p^{2}}{4m^{2}c^{2}}\right)
+m\mathcal{A}z\left(  1+\frac{p^{2}}%
{2m^{2}c^{2}}\right)
  +\mathcal{O}\left(  {1\over c^4}\right)  . \label{ham2}%
\end{equation}

This Hamiltonian can be applied to describe a system of particles
with respect to a system of reference moving with acceleration
$\mathcal{A}$, up to order $1/c^{2}$. For an accelerated nucleus the
obvious choice is the cm system of the nucleus. But then the term
carrying the acceleration correction  averages out to zero in the
center of mass, as one has ($\sum_{i}m_{i}\mathcal{A}z_{i}=0$).
There is an additional small contribution of the acceleration due to
the term proportional to $p^2$. Instead of exploring the physics of
this term, one has to account for one more correction as explained
below.

The above derivation of the Hamiltonian for particles in accelerated
frames does not take into account that the definition of the cm of a
collection of particles is also modified by relativity. This is not
a simple task as might seem at first look. There is no consensus in
the literature about the definition of the cm of a system of
relativistic particles. The obvious reason is the role of
simultaneity and retardation. Ref. \cite{Pryce48} examines several
possibilities. For a system of particles it is found convenient to
define the coordinates $q^{\mu}$ of the center of mass as the mean
of coordinates of all particles weighted with their dynamical masses
(energies). The relativistic (covariant) generalization of center of
mass is such that the coordinates $q^{\mu}$ must satisfy the
relation~\cite{Pryce48}
\begin{equation}
P^{0}q^{\mu}=%
{\displaystyle\sum\limits_{i}}
p_{i}^{0}z_{i}^{\mu}, \label{cm1}%
\end{equation}
where the coordinates of the \textit{i}th particle with respect to
the center of mass are denoted by $z_{i}^{\mu}$ and the total
momentum vector by
$P^{\mu}=%
{\displaystyle\sum\limits_{i}} p_{i}^{\mu}$. Ref. \cite{Pryce48}
chooses eq. (\ref{cm1}) as the one that is most qualified to
represent the definition of cm of a relativistic system, which also
reduces to the non-relativistic definition of the center of mass. We
did not find a better discussion of this in the literature and we
could also not find a better way to improve on this definition.

The above definition, Eq.~(\ref{cm1}), leads to the compact form, to
order $1/c^{2}$,
\begin{eqnarray}
{\displaystyle\sum\limits_{i}}
\frac{m_{i}\mathbf{r}_{i}}{\sqrt{g_{00}-\frac{v_{i}^{2}}{c^{2}}}}&=&
{\displaystyle\sum\limits_{i}} m_{i}\mathbf{r}_{i}\left(
1+\frac{v_{i}^{2}}{2c^{2}}-\frac{z_{i}\mathcal{A}
}{c^{2}}+\mathcal{O}({1\over c^4})\right)\nonumber\\ &=&0,
\label{cm2}
\end{eqnarray}
where $\mathbf{r}_{i}=(x_{i},y_i,z_{i})$ is the coordinate and
$v_{i}$ is the velocity of the {\it i}th particle with respect to
the cm.

For a system of non-interacting particles the condition in
Eq.~(\ref{cm2})
implies that, along the direction of motion,%
\begin{equation}%
{\displaystyle\sum\limits_{i}}
\mathcal{A}m_{i}z_{i}=-%
{\displaystyle\sum\limits_{i}}
\mathcal{A}m_{i}z_{i}\left(  \frac{v_{i}^{2}}{2c^{2}}-\frac{z_{i}\mathcal{A}%
}{c^{2}}\right)  . \label{cm3}%
\end{equation}
Hence, the Hamiltonian of Eq.~(\ref{ham2}) for a collection of
particles becomes
\begin{equation}
H    =%
{\displaystyle\sum\limits_{i}}
\frac{p_{i}^{2}}{2m_{i}}\left(  1-\frac{p_{i}^{2}}{4m_{i}^{2}c^{2}}%
\right)  +\frac{\mathcal{A}^{2}}{c^{2}}%
{\displaystyle\sum\limits_{i}} m_{i}z_{i}^{2}
  +U\left(  r_{i}\right) +\mathcal{O}({1\over c^{4}})  , \label{hinert}%
\end{equation}
where we have added a scalar potential $U\left( r_{i}\right)$, which
would represent a (central) potential within an atom, a nucleus, or
any other many-body system.

Notice that the term proportional to $-m\mathcal{A}z$ completely
disappears from the Hamiltonian after the relativistic treatment of
the cm. This was also shown in Ref. \cite{fewell}. It is important
to realize that non-inertial effects will also carry modifications
on the interaction between the particles. For example, if the
particles are charged, there will be relativistic corrections
(magnetic interactions) which need to be added to the scalar
potential $U\left( r_{i}\right)  =\sum_{j\neq
i}Q_{i}Q_{j}/\left\vert \mathbf{r}_{i}-\mathbf{r}_{j}\right\vert $.
As shown in Ref. \cite{fewell}, the full treatment of non-inertial
effects together with relativistic corrections will introduce
additional terms proportional to $\mathcal{A}$ and $\mathcal{A}^{2}$
in the Hamiltonian of Eq.~(\ref{hinert}), to order $1/c^{2}$. Thus,
a more detailed account of non-inertial corrections of a many-body
system requires the inclusion of $\mathcal{A}$-corrections in the
interaction terms, too. We refer the reader to Ref. \cite{fewell}
where this is discussed in more details. Here we will only consider
the consequences of the acceleration correction term in
Eq.~(\ref{hinert}),
\begin{equation}
H_{nin}=\frac{\mathcal{A}^{2}}{c^{2}}%
{\displaystyle\sum\limits_{i}} m_{i}z_{i}^{2} .\label{hinert2}
\end{equation}

\section{Reactions in stars}

Nuclei interacting in a plasma or undergoing pycnonuclear reactions
in a lattice can experience different accelerations, allowing for an
immediate application of Eq.~(\ref{hinert2}). But in order to use
this equation to measure changes induced by the gravitational fields
in stars, we assume that one can replace $\mathcal{A}$ by a local
gravitational field, $g$. This assumption requires a few comments at
this point. If we consider two nuclei participating in a nuclear
reaction in a star, they are, most likely, in a gaseous phase
following inertial trajectories in between collisions. The effect of
gravity is to modify slightly the inertial trajectories of the two
nuclei due to the difference in the gravitational field strength in
their initial and final positions. Thus the best way to study the
reaction problem is to calculate reaction rates in terms of a local
metric at a point within the star. This metric can be deduced from
General Relativity at the reaction observation point. To first-order
one can also use Eq.~(\ref{metr1}), which is shown in Ref.
\cite{Landau} to describe particles in a gravitational field. Here
instead, we will adopt the Hamiltonian of Eq.~(\ref{hinert}) as
representative of the same problem. Here we will not attempt to
prove the equality between the two procedures, and several other
issues (e.g., time-dependence of accelerations, modification of
interactions in presence of a gravitational field, etc.), leaving
this for future studies.  Our goal here is to estimate the magnitude
of the gravitational field which could produce sizable
``non-inertial corrections'' and study physical cases where such
corrections might be important and could change appreciably the
reaction rates and/or the internal structure of many body systems.

\subsection{Nuclear fusion reactions}

Nuclear fusion reactions in stars proceed at low energies, e.g., of
the order of 10 KeV in our Sun~\cite{Clayton,Rolfs}. Due to the
Coulomb barrier, it is extremely difficult to measure the cross
sections for charged-particle-induced fusion reactions at laboratory
conditions. The importance of small effects such as the correction
of Eq.~(\ref{hinert2}) in treating fusion reactions is thus clear
because the Coulomb barrier penetrability depends exponentially on
any correction. To calculate the effect of the term given by
Eq.~(\ref{hinert2}) we use,  for simplicity, the WKB penetrability
factor
\begin{equation}
P(E)=\exp\left[  -\frac{2}{\hbar}\int_{R_{N}}^{R_{C}}dr\left\vert
p(r)\right\vert \right]  , \label{pe}
\end{equation}
where $p(r)$ is the (imaginary) particle momentum inside the
repulsive barrier. The corrected fusion reaction is given by
\begin{equation}
\sigma=\sigma_{C}\cdot\mathcal{R}, \label{reduc}
\end{equation}
where $\sigma_{C}$\ is the Coulomb repulsion cross section and
$\mathcal{R}=$ $P_{corr}(E)/P(E)$ is the correction due to
Eq.~(\ref{hinert2}). The non-inertial effect is calculated using
$\left\vert p(r)\right\vert =\sqrt{2m\left[  V_{C}(r)-E\right]  }$
and
\begin{equation}
\left\vert
p_{corr}(r)\right\vert =\sqrt{2m\left[  V_{C}(r)+{\mathcal{A}^{2}%
mr^{2}\left\langle \cos^{2}\theta\right\rangle \over c^{2}}-E\right]
}
\end{equation}
where $\left\langle \cos^{2}\theta\right\rangle =1/2$ averages over
orientation and the Coulomb potential is given by
$V_{C}=Z_{1}Z_{2}e^{2}/r$. In order to assess the magnitude of the
acceleration $\mathcal{A}$ for which its effect is noticeable, we
consider a proton fusion reaction with a  $Z=17$ nucleus (chlorine)
at $E=0.1$ MeV. This is a typical fusion
reaction in stellar sites of interest. For this energy, we get $R_{C}=Z_{1}Z_{2}%
e^{2}/E=245$ fm and take $R_{N}=3.2$ fm.

\begin{figure}[ptb]
\begin{center}
\includegraphics[height=2.2477in,width=3.1375in]{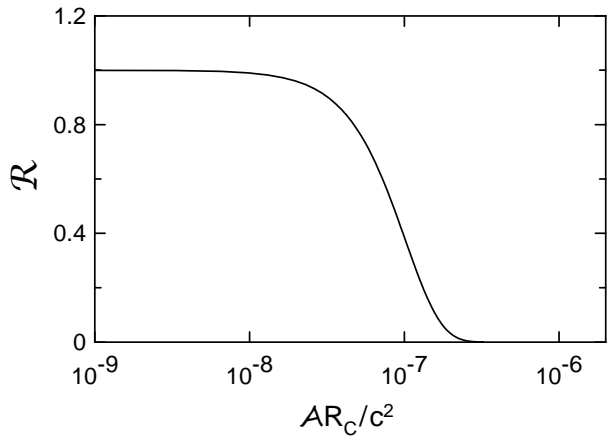}
\caption{Suppression factor due to the non-inertial effects,
$\mathcal{R}$, for fusion reactions of protons on chlorine at
$E=0.1$ MeV, and as a function of the gravitational field (in
dimensionless units).}
\label{fusion}%
\end{center}
\end{figure}

As we see in Fig.~\ref{fusion} the effect of acceleration becomes
visible for accelerations of the order
$g=\mathcal{A}=10^{-7}c^{2}/R_{C}\approx 4\times10^{27}$ m/s$^{2}$,
which is about $26$ orders of magnitude larger than the acceleration
due to gravity on Earth's surface and $15$ orders of magnitude
larger than the one at the surface of a neutron star (assuming
$M_{ns} = M_{\odot}$ and $R_{ns}=10$ km). It appears that the effect
is extremely small in stellar environments of astrophysical interest
where nuclear fusion reactions play a role. Such large gravitational
fields would only be present in the neighborhood of a black-hole.
Under such extreme conditions nuclei are likely to disassemble, as
any other structure will.

\subsection{Atomic transitions}

As an example in atomic physics, we consider the energy of the
2p$_{1/2}$ level in hydrogen which plays an important role in the
Lamb shift and probes the depths of our understanding of
electromagnetic theory. We calculate the energy shift of the
2p$_{1/2}$ level within the first-order perturbation theory and we
get
\begin{equation} \Delta E_{nin}^{2p_{1/2}}=\left\langle
2p_{1/2}\left\vert H_{nin}\right\vert 2p_{1/2}\right\rangle
={24a_{H}^{2}\mathcal{A}^{2}m_{e}\over c^{2}},
\end{equation}
where $a_{H}=\hbar^{2}/m_{e}e^{2}=0.529\ $\AA . One should compare
this value with the Lamb splitting which makes the 2p$_{1/2}$ state
slightly lower than the 2s$_{1/2}$ state by $\Delta
E_{Lamb}=4.372\times10^{-6}$ eV. One gets $\Delta
E_{nin}^{2p_{1/2}}\simeq\Delta E_{Lamb}$ for $\mathcal{A}\simeq
10^{21}$ m/s$^{2}$, which is 9 orders of magnitude larger than
gravity at the surface of a neutron star. Thus, even for tiny
effects in atomic systems, the effect would only be noticeable for
situations in which electrons are bound in atoms.

\section{Reactions in the laboratory}

The logical conclusion from the last section is that it is very
unlikely that non-inertial effects due to gravitational fields are
of relevance in stars. Nowhere, except in the vicinity of a
black-hole, accelerations are of order of $\sim 10^{20}$ m/s$^{2}$,
which would make the effect noticeable. However, there is another
way to achieve such large accelerations and that is nothing else but
the huge accelerations which occur {\it during} nuclear reactions.
For example, for a nuclear fusion reaction, at the Coulomb radius
(distance of closest approach, $R_C$) the acceleration is given by
\begin{equation}
\mathcal{A}_{C}={Z_{1}Z_{2}e^{2}\over R_{C}^{2}m_{0}}, \label{acc}
\end{equation}
where
$m_{0}=m_{N}A_{1}A_{2}/\left(  A_{1}+A_{2}\right)$ is the reduced
mass of the system and $m_{N}$ is the nucleon mass. For typical
values, $E=1$ MeV, $Z_{1}=Z_{2}=10$, and $A_{1}=A_{2}=20$, one obtains
$R_C=Z_1Z_2e^2/E=144$ fm and $\mathcal{A}%
_{C}=6.2\times10^{25}$ m/s$^{2}$. This is the acceleration that the
cm of each nucleus would have with respect to the laboratory system.

As we discussed in the introduction, the cm of the excited nucleus
is the natural choice for the reference frame. This is because it is
easier to adopt a description of atomic and nuclear properties in
the cm frame of reference. Instead, one could also chose the cm of
the colliding particles. This later (inertial) system makes it
harder to access the acceleration effects, as one would have to
boost the wave functions to an accelerated system, after calculating
it in the inertial frame. This is a more difficult task. Therefore
we adopt the cm reference frame of the excited nucleus, using the
Hamiltonian of section II. This Hamiltonian was deduced for a
constant acceleration. If the acceleration is time-dependent, the
metric of Eq.~(\ref{metr1}) also changes. Thus, in the best case
scenario, the Hamiltonian of Eq.~(\ref{hinert}) can be justified in
an adiabatic situation in which the relative velocity between the
many-body systems is much smaller than the velocity of their
constituent particles with respect to their individual center of
masses. If we accept this procedure, we can study the effects of
accelerated frames on the energy shift of states close to threshold,
as well as on the energy location of low-lying resonances.

\subsection{Reactions involving halo nuclei}

The nuclear wave-function of a (s-wave) loosely-bound, or ``halo",
state can be conveniently parameterized by
\begin{equation}
\Psi\simeq\sqrt{\alpha\over 2\pi} \ {\exp\left(  -\alpha r\right)
\over r}, \end{equation} where the variable $\alpha$ is related to
the nucleon separation energy through
$S=\hbar^{2}\alpha^{2}/2m_{N}$. In first order perturbation theory
the energy shift of a halo state will be given by
\begin{equation}
\Delta E_{nin}^{N}=\left\langle \Psi\left\vert H_{nin}\right\vert
\Psi\right\rangle={1 \over 8S} {\left( Z_{1}Z_{2}e^{2}\hbar\over
R_{C}^{2}m_{0}c\right) ^{2}}.
\end{equation}
Assuming a small separation energy $S=100$ keV, and using the same
numbers in the paragraph after Eq.~(\ref{acc}), we get $\Delta
E_{nin}^{N}= 0.024$ eV, which is very small, except for states very
close to the nuclear threshold, i.e. for $S\rightarrow 0$. But the
effect increases with $Z^2$ for symmetric systems (i.e.
$Z_1=Z_2=A_1/2$). It is thus of the order of $\Delta E_{nin}^{N}=
1-10$ eV for larger nuclear systems.

There might exist situations where this effect could be present. For
instance, the triple-alpha reaction which bridges the mass = 8 gap
and forms carbon nuclei in stars relies on the lifetime of only
10$^{-17}$ s of $^{8}$Be nuclei. It is during this time that another
alpha-particle meets $^{8}$Be nuclei in stars leading to the
formation of carbon nuclei. This lifetime corresponds to an energy
width of only $5.57\pm0.25$ eV~\cite{Ti04}. As the third alpha
particle approaches $^{8}$Be, the effects of linear acceleration
will be felt in the reference frame of $^8$Be. This will likely
broaden the width of the $^{8}$Be resonance (which peaks at
$E_{R}=91.84\pm0.04$ KeV) and consequently its lifetime. However,
this line of thought could be wrong if one assumes that the third
alpha particle interacts individually with each of the two alpha
particles inside $^8$Be, and that the effects of acceleration
internal to the $^8$Be nucleus arise from the different distances
(and thus accelerations) between the third alpha and each of the
first two. To our knowledge, this effect has not been discussed
elsewhere and perhaps deserves further investigation, if not for
this particular reaction maybe for other reactions of astrophysical
interest involving very shallow nuclear states.

\subsection{Nuclear transitions}

Many reactions of astrophysical interest are deduced from
experimental data on nucleus-nucleus scattering. Important
information on the position and widths of resonances, spectroscopic
factors, and numerous other quantities needed as an input for
reaction network calculations in stellar modeling are obtained by
the means of nuclear spectroscopy using nuclear collisions in the
laboratory. During the collision the nuclei undergo huge
acceleration, of the order of $\mathcal{A}\simeq10^{28}$ m/s$^{2}$.
Hence, non-inertial effects will be definitely important.

A simple proof of the statements above can be obtained by studying
the Coulomb excitation. The simplest treatment that one can use in
the problem is  a semi-classical calculation. The probability of
exciting the nucleus to a state $f$ from an initial state $i$  is
given by
\begin{equation}
a_{if}=-{i\over\hbar}\int V_{if}\,e^{i\omega t}dt,
\end{equation}
where $\omega=(E_f-E_i)/\hbar $, is the probability amplitude that
there will be a transition i$\rightarrow$f. The matrix element
$V_{if}=\int\Psi^*_fV\Psi_i\,d\tau$ contains a potential $V $ of
interaction between the nuclei. The square of $a_{if} $ measures the
transition probability from $i$ to $f$ and this probability should
be integrated along the trajectory.

A simple estimate could be obtained in the case of the excitation of
a initial, $J=0 $, state of a deformed nucleus to an excited state
with $J=2 $ as a result of a head on collision with scattering angle
of $\theta=180^\circ $. The perturbation $V$ is due to the
interaction of the charge $Z_1e$ of the projectile (one of the two
nuclei) with the quadrupole moment of the target (of the other)
nucleus. This quadrupole moment should  work as an operator that
acts between the initial and final states. One finds that
$V=Z_1e^2Q_{if}/2 r^3$, with
\begin{equation}
Q_{if}=e_i^2\left\langle \Psi^*_f\left\vert
3z^2-r^2\right\vert\Psi_i\right\rangle \simeq e_i^2\left\langle
\Psi^*_f\left\vert z^2\right\vert\Psi_i\right\rangle ,
\end{equation}
where $e_i$ is the effective charge of the transition.

The amplitude is then written as
\begin{equation}
a_{if}={Z_1e^2Q_{if}\over 2i\hbar}\int{e^{i\omega t}\over r^3}\,dt.
\label{ac}
\end{equation}

At  $\theta=180^\circ $  the separation $r $, the velocity $v $, the
initial velocity $v_0 $ and the distance of closest approach $s $,
are related by $v={dr/ dt}=\pm_0 v_0\bigl(1-{s/ r}\bigr),$ which is
obtained from energy conservation. Furthermore, if the excitation
energy is small, we can assume that the factor $e^{i\omega t} $ in
Eq. (\ref{ac}) does not vary much during the time that the
projectile is close to the nucleus. Then the remaining integral is
easily solved by substitution and one gets \begin{equation}
a_{if}={4Z_1e^2Q_{if}\over 3i\hbar v_0s^2}.
\end{equation}

Following the same procedure as above, we can calculate the
contribution of the Hamiltonian of Eq.~(\ref{hinert2}). In this
case, ${\cal A}=Z_1Z_2e^2/m_0r^2$ and the equivalent potential $V$
is given by
\begin{equation}
V_{nin}=\left({Z_1Z_2e^2\over m_0}\right)^2{Xm_N\over c^2r^4},
\end{equation}
where we assume that $X$ nucleons participates in the transition.
One then finds  \begin{equation} a_{if}^{nin}=\left({Z_1Z_2e^2 \over
m_0}\right)^2{32Xm_NQ_{if}\over 15is^3\hbar v_0c^2} . \end{equation}
The ratio between the two transition probabilities is
\begin{equation}
\left\vert {a_{if}^{nin} \over a_{if}}
 \right\vert^2 = \left(
8Xm_N  Z_1Z_2^2e^2\over 5s m_0^2c^2\right)^2 \label{ratV}.
\end{equation}

Applying eq. \ref{ratV} to the lead-lead collision at 500 MeV, as
mentioned in the introduction, we find $\left\vert {a_{if}^{nin} /
a_{if}}
 \right\vert^2 = (0.0093X)^2$.  This yields very small results for
the relative importance of non-inertial effects in single particle
transitions ($X\simeq 1$), but can become appreciable for the
excitation of collective states such as the giant resonances, for
which $X\gg 1$. This result is intriguing to say the least. We think
that it deserves more studies, assuming that the physics of
non-inertial effects described in section II is right. We have made
a preliminary study of theses effects in the excitation of giant
resonances in relativistic heavy ion collisions using Eq.
(\ref{hinert2}) which seem to confirm this statement.

\subsection{Electron screening of fusion reactions}

In laboratory measurements of nuclear fusion reactions one has found
enhancements of the cross sections due to the presence of atomic
electrons. This screening effect leads to an enhancement in the
astrophysical S-factor, or cross section:
\begin{equation}
S_{lab}\left( E\right) =f\left( E\right)  S\left( E\right)
=\exp\left[ {\pi\eta\Delta E\over E}\right]  S\left( E\right),
\end{equation}
where $\eta\left( E\right) =Z_{1}Z_{2}e^{2}/\hbar v$, and $v$ is the
relative velocity between the nuclei. The energy $\Delta E$ is equal
to the difference between the electron binding energies in the
$\left(Z_{1} + Z_{2}\right) $-system and in the target atom
($Z_{2}$). For light nuclei it is of the order of 100 eV,  enhancing
fusion cross sections even for fusion energies of the order of 100
KeV. For more details we refer the interested reader to
Ref.~\cite{Ro01}.

An intriguing fact is that this simple estimate, which is an upper
value for $\Delta E$, fails to reproduce the experimental data for a
series of cases. In Ref.~\cite{Ba97}\ several small effects, ranging
from vacuum polarization to the emission of radiation, have been
considered but they cannot explain the experimental data puzzle.
Besides vacuum polarization, atomic polarization is one of the
largest effects to be considered (among all other small
effects~\cite{Ba97}).

Non-inertial corrections contribute to polarization potential
\begin{equation}
V_{pol}=-\sum_{n\neq0}{\left\vert \left\langle 0\left\vert
H_{nin}\right\vert n\right\rangle \right\vert ^{2}\over
E_{n}-E_{0}}.
\end{equation}
An estimate based on hydrogenic wave functions for the atom yields
\begin{equation}
V_{pol}\left(  r\right)  \simeq-\frac{1}{E_{n0}}\left( \frac{
Z_{1}Z_{2}e\hbar}{m_{0}c}\right) ^{4}\frac{\exp\left( -2\alpha
r\right) }{r^{4}}.\label{vpol}
\end{equation}
Assuming $\alpha\cong1/a_{H},$ $E_{n0}=E_{n}-E_{0}\cong10$ eV and
using Eqs.~(\ref{pe}) and (\ref{reduc}) to calculate the
modification of the fusion cross sections due to this effect, we
find the cross section for D(d,p)T and $^{6}$Li$\left( d,\alpha
\right) ^{4}$He can increase by up to 10\%. This is surprising
compared with the smaller values reported on Table 1 of Ref.
~\cite{Ba97}. It is not a very accurate calculation as it relies on
many approximations. But it hints for a possible explanation of the
difference between the experimental and theoretical values of
$\Delta E$, as discussed in Ref.~\cite{Ro01}.

In stars, reactions occur within a medium rich in free electrons.
The influence of dynamic effects of these electrons was first
mentioned in Ref.~\cite{Mi77} and studied in Ref.~\cite{Car88}. The
underlying assumption is that the Debye-Hueckel approximation, based
on a static charged cloud, does not apply for fast moving nuclei. In
fact, most of the nuclear fusion reactions occur in the tail of the
Maxwell-Boltzmann distribution. For these nuclear velocities
Ref.~\cite{Car88} finds that an appreciable modification of the
Debye-Hueckel theory is necessary. One has to add to this finding
the fact that the nuclei get very strongly accelerated as they
approach each other, and this increases further the deformation of
the Debye-Hueckel cloud.

\section{Conclusions}

In summary,  assuming that the Hamiltonian for a system of particles
moving in an accelerated frame contains a correction term of the
form given by Eq.~(\ref{hinert2}), we have explored the non-inertial
effects for a limited set of nuclear reactions in stars and in the
laboratory. These results are somewhat surprising and present a
challenge to our understanding of accelerated many-body systems.

In the case of stellar environments, we have shown that only in the
neighborhood of black-holes would non-inertial effects become
relevant. But then the whole method adopted here is probably not
rigorous enough, as one may have to use the full machinery of
general relativity. Nonetheless, it is very unlikely (and perhaps
unimportant, except maybe for science-fiction-like time-traveling)
that internal structures of any object is of any relevance when it
is extremely close to a black-hole.

The apparent reason for the appearance of non-inertial effects in
many-body systems is that the non-inertial term of
Eq.~(\ref{hinert2}) only appears when relativistic corrections are
included, what has precluded its consideration in previous studies,
especially for reactions that are thought to be fully
non-relativistic such as fusion reactions in stars. The main
question is whether the relativistic definition of the center of
mass, through Eq. (\ref{cm1}) as proposed by Pryce in Ref.
\cite{Pryce48} contains the right virtue of describing correctly the
center of mass frame of relativistic many-body systems.

Even in the case of high energies nuclear collisions the intrinsic
structure of the nuclei are sometimes an important part of the
process under study. Fictitious forces will appear in this system
which might not average out and appreciably influence the structure
or transition under consideration. It is surprising that, for a
reason not quite understood, this effect has been overseen in the
literature so far.

{\bf Acnowledgments}

This work was partially supported  by the U.S. DOE grants
DE-FG02-08ER41533 and DE-FC02-07ER41457 (UNEDF, SciDAC-2), the NSF
grant PHY0652548, and the Research Corporation under Award No. 7123.


\begin{thebibliography}{99}

\bibitem {Landau}L.D. Landau and E.M. Lifshitz, Classical theory of
fields (Second Edition), Pergamon Press, p.286

\bibitem {Pryce48}M. H. L. Pryce, Proc. Roy. Soc. A195, 62 (1948).

\bibitem{fewell} M.P. Fewell, Nucl. Phys. {\bf A425} (1984) 373.

\bibitem {Clayton}D.D. Clayton, Principles of Stellar Evolution and
Nucleosynthesis, McGraw-Hill, New York, 1968.

\bibitem {Rolfs}C. Rolfs, W.S. Rodney, Cauldrons in the Cosmos, University of
Chicago Press, Chicago, IL, 1988.

\bibitem {Ti04}D. R. Tilley et al., Nucl. Phys. A745, 155 (2004).

\bibitem {Ro01}C. Rolfs, Prog. Part. Nucl. Phys. 46, 23 (2001).

\bibitem {Ba97}A.B. Balantekin, C.A. Bertulani and M.S. Hussein, Nucl. Phys. A
627, 324  (1997).

\bibitem {Mi77}H.E. Mitler, Ap. J.  212, 513 (1977).

\bibitem {Car88}C. Carraro, A. Sch\"{a}ffer and S.E. Koonin,  Ap. J. 331, 565
(1988).

\end{thebibliography}
\end{document}